# A Wireless, Inexpensive Optical Tracker for the CAVE™

Ehud Sharlin        Pablo Figueroa        Mark Green        Benjamin Watson

*Computer Graphics Laboratory, Department of Computing Science, University of Alberta*
*{ehud,pfiguero,mark,watsonb}@cs.ualberta.ca*

## Abstract

*CAVEÔ displays offer many advantages over other virtual reality (VR) displays, including a large, unencumbering viewing space. Unfortunately, the typical tracking subsystems used with CAVEÔ displays tether the user and lessen this advantage. We have designed a simple, low-cost feet tracker that is wireless, leaving the user free to move. The tracker can be assembled for less than $200 US, and achieves an accuracy of ±10 cm at a 20 Hz sampling rate. We have tested the prototype with two applications: a visualization supporting close visual inspection, and a walkthrough of the campus. Although the tracking was convincing, it was clear that the tracker's limitations make it less than ideal for applications requiring precise visual inspection. However, the freedom of motion allowed by the tracker was a compelling supplement to our campus walkthrough, allowing users to stroll and look around corners.*

## 1. Introduction

The CAVE™ display [6] is now widely used in virtual reality (VR) systems to visualize complex scientific data, prototype complex industrial designs, and simulate realistic environments for training. It has many advantages over alternatives such as the head-mounted display, including high resolution and large field of view (FOV). In addition, because there is no need for occluding and often awkward headgear, users enjoy greater freedom of motion and experience less fatigue.

Unfortunately, the need to track the user's head position often reduces this latter advantage (Figure 1). The most commonly used tracking solution, electromagnetic trackers, tether users with cabling attached to the head and other parts of the body. Usoh et al [22] have identified this as a significant problem. Other tracking technologies [8,11,24] are quite expensive, still experimental, or restrict freedom of motion.

The cost of CAVE™ systems (often in the range of $500,000 US) has also limited their use. Relative to these prices, the expense of tracking technology was a minor consideration. However, recent changes in PC technology have enabled the construction of more affordable CAVE™ facilities [12], making the cost of the tracking subsystem a more pressing concern.

We are applying the VizRoom, the University of Alberta's high-end CAVE™ facility, towards wayfinding training and research. Researchers will gain control of experimental variables and measures not available in the real world. Users will be able to explore hostile or distant environments, and gain a working knowledge of them much more quickly than normally possible. For these sorts of applications, the restricted freedom of motion resulting from the use of typical electromagnetic trackers is a particularly onerous problem. We are also experimenting with a low-cost, PC-based CAVE™ display, which we call the Cavelet. The need for a wireless, low-cost alternative to existing tracking technologies is clear.

Our solution is an optical tracking system based on structured light (SL) technology [2,3,7,10,14,16,18,23]. A commodity laser projects a line on the VizRoom center screen, a few inches above the floor. Using a simple video camera and card, we compare this line to the reflections of the laser off the user's legs, making it possible to calculate the user's position on the floor (Figures 1 & 2). Users are completely untethered and need not wear any special equipment. The cost of the system (excepting the PC and camera) is less than $200 US.

There are limitations to this technology. By tracking the feet instead of the head with only two degrees of freedom (DOFs), we introduce inaccuracy into the display. This is particularly obvious when objects are close to the user. However, for many applications like the wayfinding research described above, objects are at a distance that does not require much precision, and the new wireless tracking system is quite effective.

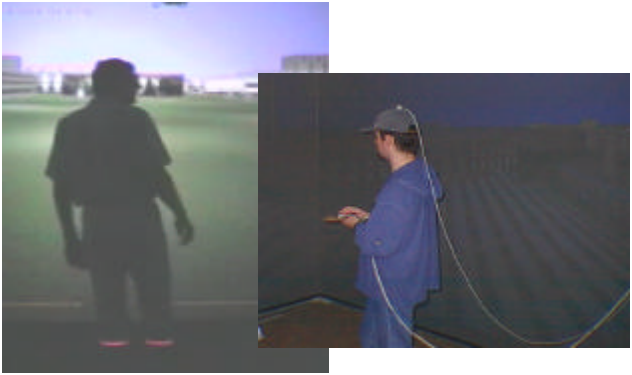

**Figure 1. Untethered optical tracking vs. tethered electromagnetic tracking.**

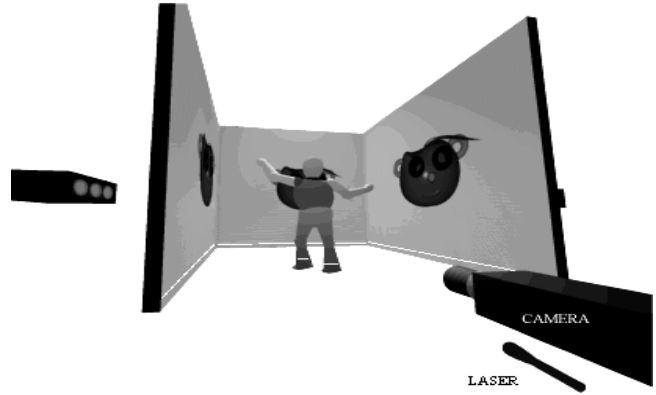

**Figure 2. The optical tracker in the VizRoom - an overview**

In the remainder of this paper, we describe this new tracking system and our initial experience with it. In sections 2 and 3, we discuss the characteristics of the CAVE™ technology that makes it amenable to limited DOF tracking. Sections 4, 5 and 6 review related optical tracking methods and structured light research. Section 7 outlines our basic concept. Sections 8, 9 and 10 detail the geometry model, hardware, and software used in the tracker. We describe the tracker's accuracy and initial user impressions in Section 11. Sections 12 and 13 conclude the paper and describe future research.

## 2. CAVE™ sensitivity to tracking errors

The CAVE™, unlike the common head-mounted display (HMD), has static projection planes. This makes the CAVE™ much less sensitive to head rotation tracking errors than devices in which the projection planes move with the user's head [6]. Ideally, perfect rotation about the projection point in the eye does not create any changes in the projected scene (with the two eye points of stereoscopic display, this ideal can never be achieved). As a room size, static projection plane system, the CAVE™ is also much less sensitive to combined tracking errors (position and orientation) than HMDs [6]. In addition, the CAVE™'s relative sensitivity to errors decreases as the distance to the projected objects increases.

## 3. Trading DOFs for freedom of motion

Given that CAVEÔ displays have less sensitivity to tracking error, we thought it worthwhile to investigate whether real freedom of motion might compensate for some tracking inaccuracy. We believe there is a large class of applications in which freedom of motion is a more pressing concern than tracking precision.

To investigate this question, we designed a simple two dimensional position tracker using a basic SL technique. Because we ignore orientation, we lose some accuracy in stereoscopic projection. However, for the reasons outlined above, this may be a minor concern. We also are forced to assume that the user's head is at a constant height. In return, users gain great freedom of motion, and the cost of tracking drops steeply. Below we describe the tracker, and our initial experience with it.

## 4. Machine vision and structured light

Current imaging and electro-optics technology offers many off-the-shelf machine vision solutions to three-dimensional object tracking. Machine vision techniques can be classified into two main categories: passive and active techniques. Passive techniques are closely related to the way humans perceive the world. Some of these techniques are relatively straightforward and suffer from inherent problems of information loss while transferring a 3D scene onto a 2D image [18]. Active machine vision techniques take a different approach (not at all like the human natural visual system) to depth extraction. These techniques are based on illumination of the scene with light (with defined qualities) and depth extraction from the reflection. Common active techniques are laser radar and structured light ranging [7,18]. Laser radar techniques can be very accurate but are generally very complex and costly systems. SL techniques are much less complex and are well established in industrial computer vision applications. SL is defined as: "…the process of illuminating an object (from a known angle) with a specific light pattern. Observing the lateral position of the image can be useful in determining depth information…" [14]. Inspired by "The Office of the Future" [19] (see section 6), we have chosen a simple active machine vision SL technique for our VizRoom tracker.

## 5. Optical tracking in virtual reality systems

Virtual Reality tracking technology includes mechanical, ultra-sonic and optical systems [11]. However, the dominant trackers are still the electromagnetic trackers mentioned previously. In [4], Chance et al. describe a wide area tracking system in 3 DOFs based on stereo imagery, with two CCD (Charged Coupled Device) cameras and a light bulb on the user's backpack. In [24] Welch et. al describe the development of an 6 DOF optical tracker called HiBall, for HMDs. The system was motivated by the need to support real walking as an interaction technique in a virtual walkthrough. The system consists of an array of light emitting diodes (LEDs) mounted on the room's ceiling. On top of the HMD, a small cluster of 6 photodiodes senses the light emitted from the ceiling's LEDs. The 6 DOFs of the HMD are extracted from the known geometry of the LED patterns and sensors. By adding more LED ceiling panels the system supports wide area tracking. The HiBall's goals are much broader than our goals and its solution is much more complex and specialized than what we require.

## 6. The "Office of the Future" and imperceptible structured light

Our work was inspired by the "Office of the Future", presented by Rasker, Welch, Cutts, Lake, Stesin and Fuchs [19]. The "Office of the Future" is a work environment that supports complete integration of physical objects and synthetic virtual objects. Computer vision techniques provide a means of tracking the physical objects in the environment. An active vision – SL technique is used to extract object positions. The SL pattern projected on to the environment is a vertical binary bar grid. However, though working in the visible spectrum, the authors do not want the user to perceive the projected pattern. They suggest the use of imperceptible SL produced by high frequency, time-multiplexed projection of the binary light patterns and their complement. At high temporal frequencies, human vision is not sensitive to this repetition (while a synchronized camera can still extract it). Again, the goals of the "Office of the Future" are much broader than ours. We chose a much simpler SL technique for our tracker.

## 7. Design goals and fundamental concepts

Our optical tracker's main design goals were:

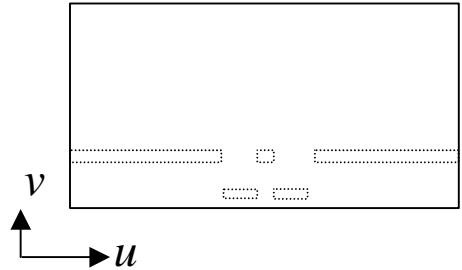

**Figure 3. Laser line reflection from back wall and user's feet as seen in image coordinates. The vertical disparity in the imaged line corresponds to the physical distance from the user to the camera.**

- The tracker must not limit user movement in any way nor require the user to carry any device.
- The tracker should be very simple and inexpensive.

In addition, we had many other concerns and goals. First, we wanted to ensure that the tracker's accuracy was roughly 10 cm. We also wanted to sample position as frequently as possible – at least 15 Hz. The tracker would have to function in a noisy environment, without interfering with display or stereoscopic synchronization. And finally, the tracker must be safe for users at all times.

Our tracker achieves these design goals by optically tracking the user's feet. Since the VizRoom is a confined space and the user must be connected to the ground, this simple approach seems sound and will enable a very straightforward implementation of SL technology. The SL technique we implemented is a single line projection across the VizRoom's floor, using a laser line pattern projector. An "artist view" of the tracker inside the VizRoom is presented in Figure 2: note the video camera, the laser projector mounted below the camera, the projected horizontal line across the lower part of the back wall and the reflection from the user's feet. Note also that the user is released from any wires or attachments and can walk freely in the environment.

Figure 3 presents the resulting image sensed by the camera, assuming the sensor is sensitive only to the laser line reflection. Even without elaborating on the geometry of the problem it is quite apparent that the user position on the floor can be extracted from the image. By comparing the vertical position of the reflections off the user's feet to the reflection off the back wall, we can approximate the user's depth. The horizontal position of the feet reflections corresponds to the user's lateral position.

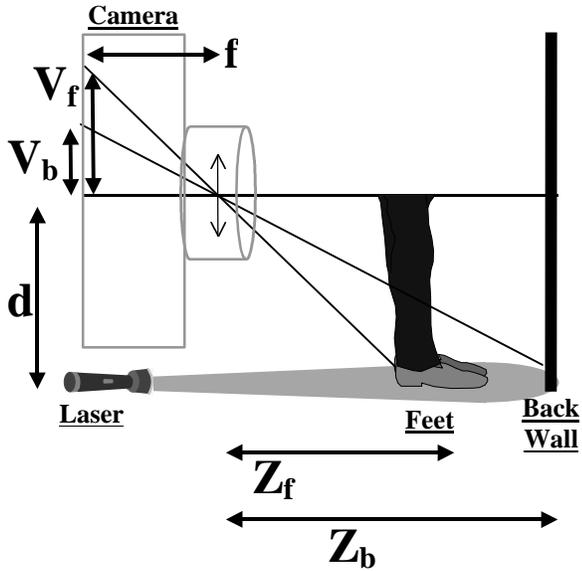

**Figure 4. Side view of tracking geometry. Focal length exaggerated for illustration purposes.**

## 8. Detailed tracking method

Figures 4 and 5 present a simplified adaptation of SL techniques to our case, which can be viewed as a simple triangulation problem. Figure 4 presents a side view of the tracking scenario while figure 5 presents a top view. The figures show the laser projection, the reflection from the back of the VizRoom and from the user's feet, and the camera's lens and image plane. The physical distances (for both figures 4 and 5, measured in cm) in the VizRoom are:

- $d$ - The vertical distance between the laser projector and the camera.
- $Z_b$ - the distance in depth between the camera and the back of the VizRoom.
- $Z_f$ - the distance in depth between the camera and the reflection from the user's feet.
- $X_f$ - the lateral distance between the center of the camera's FOV and the reflection from the user's feet.
- $f$ - The camera's focal length.

The image-plane parameters (distances for both figures 4 and 5, measured in pixels) are:

- $V_f$ - The vertical coordinate of the reflection from the user's feet.
- $V_b$ - The vertical coordinate of the reflection from the back of the VizRoom.
- $U_f$ - The horizontal coordinate of the reflection from the user's feet.

Using triangle similarity we can extract $Z_f$ from figure 4:

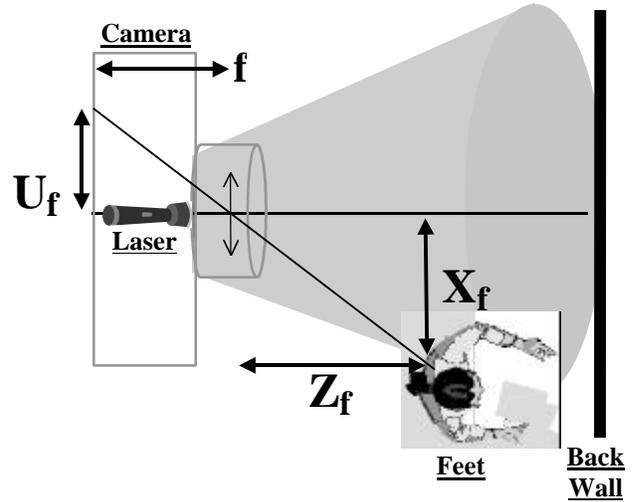

**Figure 5. Top view of the tracking geometry. Focal length exaggerated for illustration purposes.**

$$Z_f = \frac{d \cdot f \cdot Z_b}{d \cdot f + Z_b \cdot (V_f - V_b)} \quad (1)$$

And $X_f$ from figure 5:

$$X_f = \frac{U_f \cdot Z_f}{f} \quad (2)$$

Hence, given the reflection from the user's feet, we can extract the user's position in 2 DOF by a straightforward, simple algorithm.

Intuitively speaking, how much light should we expect at the sensor after our emitted laser is transmitted and reflected back to the sensor from the user's feet? A comprehensive discussion of radiometry and photometry fundamentals can be found in various sources [5,13,17]. Initially, we attempted to extract reflections using a constant luminance threshold. However, this proved inadequate. Instead, we use an adaptive threshold that varies with the expected distance to the user (we base our expectations on the vertical image coordinate). With reflections that should be closer, we use a larger threshold. We use a smaller threshold with reflections that are likely to be more distant.

## 9. Hardware implementation

Adding an optical system to the VizRoom is somewhat problematic. The VizRoom is a very noisy optical environment. First, the VizRoom projectors emit light in the entire visible spectrum. Second, the VizRoom's liquid crystal display (LCD) shutter glasses are synchronized by near infrared (IR) pulses emitted by an LED emitter. These facts impose some significant restrictions on our tracking system:

- In order to avoid any interference with the VizRoom's projected scenes we must: (1) physically avoid any interference with the projected scenes and/or (2) work in the non-visible spectrum. From a cost effectiveness point of view (2) implies working in the near IR spectrum, since far IR imaging systems are too expensive.
- In order to avoid any interference with the shutter glasses' sensors we must: (1) work outside their spectral region (which practically implies working somewhere in the visible region) and/or (2) never direct the laser beam into the sensor's FOV, in order not to block the LED's pulses.
- In order to receive a useful reflection image from the tracker's camera we should keep the S/N (signal to noise) ratio high, where the VizRoom's projected scenes are the noise. This implies (1) optically filtering the image around the monochromatic laser's wavelength and/or (2) working outside of the visible spectrum, probably in the near IR region and/or (3) utilizing image processing algorithms on the perceived image: threshold tests, edge detection, spatial filtering - looking only for horizontal (or nearly horizontal) lines in the image.

Since our proposed tracking system uses a laser, eye safety design considerations are essential. Briefly, the relevant laser hazards are categories into classes 2, 3A and 3B [14,21]. Class 2 ('Caution') lasers are low power (less than 1 mW) visible lasers. Normal exposure to these lasers does not cause any permanent damage to the retina and they are considered "eye safe" unless the eye is forced to stare into the beam. Class 2 lasers can be implemented in an application with minimal safety concerns and can be compared to the hazards of film or slide projectors [21]. Class 3A ('Danger') are lasers in the visible region with power from 1mW to 5mW. Class 3B ('Danger') are lasers in the visible region with power from 5mw to 500mw and any invisible (UV or IR) lasers with power lower than 500mw. Focusing a class 3A or 3B laser into the eye can result in eye damage.

Line projecting lasers disperse power over the projected fan angle. Working with a large fan angle (i.e.: 90 degrees) considerably reduces the actual intensity of the laser perceived by the eye, if directly hit. This fact practically upgrades most of the line lasers projectors safety category to class II. However, since some of the line projectors' emission is not uniformly distributed, a central "hot spot" may cause these lasers to be classified as class III [14]. *Because of the lack of proper power measurement equipment we chose to treat our prototype's laser as "Dangerous" during all our experiments.*

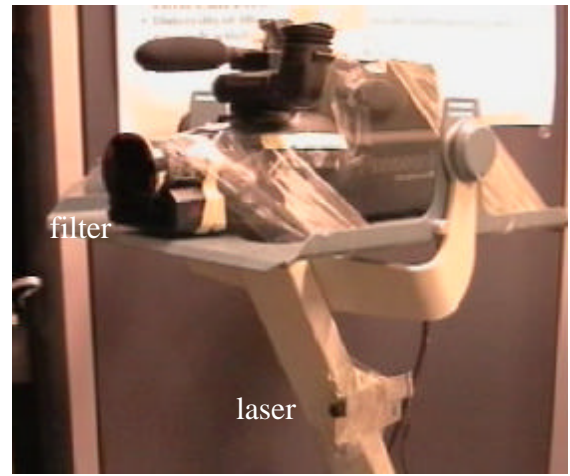

**Figure 6. Prototype hardware implementation**

Trying to combine these different considerations into a unified system might result in various solutions. We chose to combine these considerations into the following framework:

- We use a low-altitude, floor-parallel laser, solving many of the eye safety issues as well as potential coexistence problems. The low altitude laser beam does not enter the projected screen scenes, the shutter glass sensor's FOV, or the eyes' FOV.
- Using lasers in this configuration, eye safety is almost inherent. However, caution dictates that we use safest laser possible. Since these sorts of lasers emit light in the visible spectrum, we use a narrow optical filter and an image processing algorithm to maintain a reasonable S/N ratio.

The tracker prototype hardware implementation is shown in figure 6. We used a 7mw, 670 nm (red) laser diode, with a 90-degree fan angle line projector [15,26]. The inexpensive and small laser was mounted below a regular video camera. The centers of the FOVs of the laser and the camera are kept parallel. A low-priced, plastic optical long-pass filter was mounted on top of the camera's objective lens in order to reduce the sensitivity of the camera in spectral regions other than those of the laser [1]. As a "real-time image processor" we used a 166MHz Pentium MMX PC and an inexpensive WINNOV video capture board [25]. The current combined cost of the laser, video capture card and the optical filter is under 200$ US (we exclude the video camera and the PC because they are common accessories).

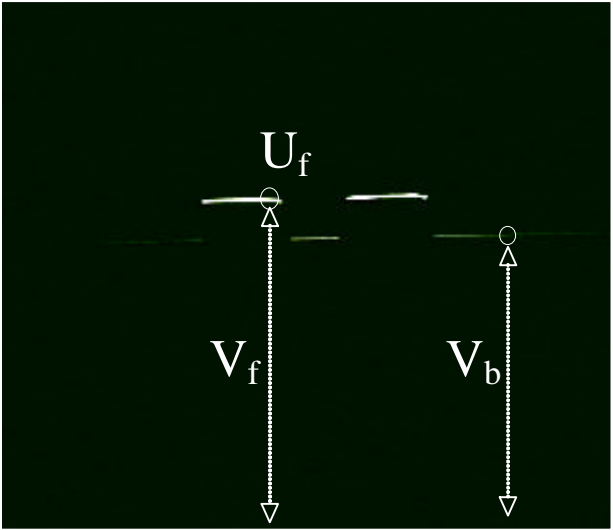

**Figure 7. A sample tracking image. The tracking values are superimposed.**

## 10. Software implementation

In order to extract $Z_f$ and $X_f$ from equations (1) and (2) the tracker has to extract, in real-time, the values of $V_f$, $V_b$ and $U_f$ from tracking images that look like the one shown in Figure 7. The video capture card we used does not have any processing power onboard, so image processing is performed on a frame to frame basis by the Pentium host processor. In order to maintain a sufficient update rate, the image-processing algorithm has to be kept minimal. Initially we had hoped to extract the different tracking values by a simple grayscale threshold test. However, the current optical filter is not narrow enough to maintain a high S/N ratio between the laser reflections (the signal) and the VizRoom scenes (the noise). In the end, we used a simple horizontal edge-detection test with an adaptive distance related threshold [9].

First, the value of $V_b$ is extracted, during the tracker's calibration process, since the position of the back screen reflection does not change as long as the tracker does not move. The software then extracts the feet position by testing every pixel value in the image (P(u,v), such that $v>V_b$) by the following test T(P(u,v)) (see equation 3). The position of the feet is chosen as the center of mass of a continuous horizontal sequence of pixels with T(P(u,v))=1 values.

$$T = sign\{P(u,v) - \frac{P(u,v+1) + P(u,v-1)}{2} - ATH(v - V_b)\} \quad (3)$$

where:

(u,v): image coordinates.
P(u,v): pixel value at image coordinates.
ATH (v- $V_b$) - Adaptive Threshold. ATH is roughly proportional to (v- $V_b$) (see equation 1).

## 11. Preliminary Results

We have performed preliminary tests comparing our tracker output to a Polhemus electromagnetic tracker output and to the physical location while performing dynamic tracking (i.e., the user is moving). Based on these tests our current prototype tracks the user's feet position in ~20Hz frame rate and with better than ±10 cm accuracy. We believe that filtering and predication will greatly improve these tracking results.

We performed initial testing of the tracker with two different types of applications. In the first, a simple visualization, a single virtual object floated in the center of the VizRoom. In the second, users explored a virtual representation of the University of Alberta campus.

In the simple visualization, the limitations of the tracker were fairly clear. When real world objects are quite close, small head motions (not requiring any movement of the feet) are an important part of any visual inspection. Our tracker does not support this behavior. Without tracking of vertical position, examining the top or bottom of the virtual object was difficult. We were pleased to find that we did not experience any diplopia, despite the complete absence of orientation tracking – although the accuracy of the represented stereoscopic depth is probably questionable. Without any cabling, it was a simple matter to walk completely around the virtual object. Nevertheless, it is clear that our tracker is not the best solution for applications requiring close visual inspection.

However, our tracker proved almost ideal for the campus walkthrough. Because the buildings in the campus are always at least a few meters away, the limitations of the tracker had minimal impact on user behavior. The absence of any cabling supported a natural sort of exploration, enabling users to stroll back and forth to look around corners. As Usoh et al [22] have emphasized, this freedom of motion is quite compelling. We expect to make heavy use of the tracker as our experiments with the application continue.

## 12. Future Work

Our current efforts are directed in four main directions:

i. *Additional DOFs*: we are already thinking about simple ways of upgrading the tracker to three or four DOFs.
ii. *Improved filtering*: we are planning to use Kalman filtering in order to improve our dynamic tracking

results. This technique is widely used in dynamic systems and was already introduced for predictive tracking in VR (see for example [24]).

iii. *User evaluation tests*: we still must formally test our claim that user interaction in many CAVE™ applications can be sufficient with only 2 (or 3) DOF tracking.

iv. *Wide area tracking*: theoretically, our tracker can be scaled up simply to large area tracking without additional hardware and without reducing sample rate.

## 13. Summary

We have described a simple, inexpensive optical tracker. The tracker is easily assembled for less than $200 US from simple commodity components. Because it does not track vertical position or any orientation, the tracker is not appropriate for precision applications requiring close visual inspection. However, there is a large class of applications (e.g. walkthroughs) in which displayed objects are more distant, and in which whole-body turning and walking must be supported. Our tracker is an inexpensive way of supporting this compelling freedom of motion.

## 14. Acknowledgments

We thank Steve Sutphen for his long hours of technical support and his wise advice. We would also like to thank Lloyd White, Dima Brodsky and Oscar Meruvia for their help during the tracker experiments, Charles Jobagy and Rick Hughes for their technical support, Paul Ferry for the Mace demo and Amira Sharlin for the 3D Rhino model of the tracker.